\documentstyle[12pt]{article}

\begin{document}
\begin{titlepage}

\hfill{OITS-532}

\hfill{UM-P-93/115}

\hfill{OZ-93/26}

\hfill{IP-ASTP-32}


\vskip 1 cm

\centerline{\Large \bf Model for a Light $Z'$ Boson}

\vskip 1.5 cm

\centerline{{\large R. Foot}$^{(a)}$,
{\large X.-G. He}\footnote{he@bovine.uoregon.edu}$^{(b)}$, {\large H.
Lew}\footnote{lew@phys.sinica.edu.tw}$^{(c)}$
{\large and
R. R. Volkas}\footnote{U6409503@hermes.ucs.unimelb.edu.au}$^{(d)}$}

\vskip 1.0 cm
\noindent
\centerline{{\it (a) Physics Department, McGill University,
3600 University Street,}}
\centerline{{\it Montreal, Quebec, H3A 2T8, Canada. }}

\vskip 0.25cm
\noindent
\centerline{{\it (b) Physics Department, ITS, University of Oregon,}}
\centerline{{\it Eugene, OR 97403-5203 U.S.A.}}

\vskip 0.25cm
\noindent
\centerline{{\it (c) Institute of Physics, Academia Sinica,}}
\centerline{{\it Nankang, Taipei, Taiwan 11529, R.O.C.}}

\vskip 0.25cm
\noindent
\centerline{{\it (d) Research Centre for High Energy Physics,
School of Physics,}}
\centerline{{\it University of Melbourne, Parkville 3052, Australia.}}

\vskip 1.0cm

\centerline{Abstract}

\noindent
A model of a light $Z'$ boson is constructed and phenomenological bounds
are derived. This $Z'$ boson arises from a very simple extension to the
Standard Model, and it is constrained to be light because the vacuum
expectation values which generate its mass also break the electroweak
gauge group. It is difficult to detect experimentally because it couples
exclusively or primarily (depending on symmetry breaking details) to
second and third generation leptons. However, if the $Z'$ boson is
sufficiently light, then there exists the possibility of the two-body
decay $\tau \rightarrow \mu Z'$ occuring. This will provide a
striking signature to test the model.

\end{titlepage}

The success of the Standard Model (SM) has led many people to believe
that it is the correct low energy theory for physics below about 100
GeV. Despite this success there are still many ways in which the SM
might be incomplete. For example, experiments may reveal neutrino masses.
Another possibility is that the gauge sector is incomplete.

The SM uses the gauge group $G_{SM} =
$\ SU(3)$_c\otimes$SU(2)$_L\otimes$U(1)$_Y$ with the fermion transformation
laws being,
\begin{eqnarray}
& Q_L \sim (3,2)(1/3),\quad u_R \sim (3,1)(4/3),\quad d_R \sim
(3,1)(-2/3),& \nonumber\\
& \ell_L \sim (1,2)(-1)\quad {\rm and}\quad \ell_R \sim
(1,1)(-2).&\
\end{eqnarray}
Before the discovery of neutral currents the theoretical need for the
U(1) factor in $G_{SM}$ was recognised since this was the minimal way to
incorporate the U(1) of electromagnetism and the SU(2) which contained
the charged current weak interactions \cite{glashow}.
[The U(1) inside the SU(2) could
not be used because it gave the wrong electric charges.] While not
strictly necessary theoretically, it might be that there are other gauged
U(1) symmetries. In other words, the gauge symmetry (below some scale)
may effectively be given by $G_{SM}\otimes$U(1)$'$. In this paper, we are
interested in examining the possibility that nature is effectively
described by a gauge theory with gauge group $G_{SM}\otimes$U(1)$'$ with
all gauge boson masses less than about 100 GeV. This is an important
question, since it would mean that low energy physics is not described
by $G_{SM}$ but rather by $G_{SM}\otimes$U(1)$'$.

How are we to choose the spectrum of U(1)$'$ charges? We impose three
constraints: First, we will assume that
the new gauge group $G_{SM}\otimes$U(1)$'$ is
anomaly-free under the condition that the standard quarks and leptons are
the only fermions in the model. To keep the fermion spectrum minimal we
will in particular exclude right-handed neutrinos. Second, the nonzero
vacuum expectation values (VEVs) which break U(1)$'$ should also break
the electroweak gauge group. This ensures that the symmetry breaking
scale for U(1)$'$ cannot be made arbitrarily high. We will also demand
that all Higgs multiplets couple to fermions through Yukawa terms. This
serves to connect the U(1)$'$ charges of the fermions with those of
the Higgs bosons. Third, we would like
the $Z'$ coupling constant to be as large as phenomenology allows. This
will maximise the testability of our model.

The condition of anomaly freedom informs us that the U(1)$'$ must couple
differently to the different generations. This is because U(1)$_Y$ is
the only generation blind symmetry that is anomaly-free with respect to
$G_{SM}$. By using
this piece of information together with the third criterion stated above
we can narrow down the choices considerably. Most experiments are
done by using the interactions of the first generation fermions, since
these comprise ordinary matter. Any $Z'$ which couples to the first
generation will be more heavily constrained than one which couples to
second and third generation fermions only. Since we are interested
in the possibility of a very light $Z'$, we thus assume that the
U(1)$'$ charges of the first generation fermions are all zero. Another
stringent constraint on the $Z'$ interactions arises from flavour-changing
neutral current (FCNC) processes. If it does not couple universally to
quarks, then in general the interactions of the $Z'$ will not conserve
flavour. This means there will be no GIM mechanism, and experimental bounds
on processes such as $K-\overline{K}$ mixing will render the $Z'$
coupling constant rather small. We are thus lead to suppose that our
$Z'$ couples only to second and third generation leptons.

So, we start by assuming the most general U(1)$'$ charge assignments
consistent with the above assumptions:
\begin{eqnarray}
& \ell_{2L} \sim (1,2)(-1,a_1),\qquad e_{2R} \sim
(1,1)(-2,b_1),&\ \nonumber\\
& \ell_{3L} \sim (1,2)(-1,a_2),\qquad e_{3R} \sim
(1,1)(-2,b_2).&\
\label{Y_prime_charges}
\end{eqnarray}
Anomaly cancellation implies that $a_1 = -a_2$, $b_1 =
-b_2$ and $a_1 = \pm b_1$. The sign ambiguity in the last
equation is of no consequence since one can be obtained from the other
be renaming $e_{2,3R}$ as $e_{3,2R}$; we choose the plus sign. Note
that the fields in Eq.~(\ref{Y_prime_charges}) are the weak
eigenstates. In general, the mass eigenstates will be linear
combinations of the weak eigenstates.

There are only three choices of weak eigenstates which have a type
of GIM mechanism.
\vskip .3cm
\noindent
(1) $\ell_{2L} = \mu_L,  \ell_{3L} = \tau_L, e_{2R} = \mu_R,
e_{3R} = \tau_R$
\vskip .3cm
\noindent
(2) $\ell_{2L} = \mu_L, \ell_{3L} = \tau_L, e_{2R} = \tau_R,
e_{3R} = \mu_R$
\vskip .3cm
\noindent
(3) $\ell_{2L} = {(\mu + \tau)_L \over \sqrt{2}},
\ell_{3L} = {(\mu - \tau)_L \over \sqrt{2}},
e_{2R} = {(\mu + \tau)_R \over \sqrt{2}},
e_{3R} = {(\mu - \tau)_R \over \sqrt{2}}.$

\vskip .3cm
\noindent
(In this equation we have denoted mass eigenstates by $\mu$ and $\tau$.)
The first case corresponds to gauged $L_{\mu} - L_{\tau}$ and has
been discussed previously \cite{Le-Lmu}.
Note that since $L_{\mu} - L_{\tau}$
is a symmetry of the standard model, this symmetry is not
broken by fermion masses (assuming the minimal fermion content
of 15 Weyl fields per generation). The second case [(2) above]
corresponds to gauged axial $L_{\mu} - L_{\tau}$.
This case has not been discussed previously as far as we are aware.
In this case, since the $\mu$ and $\tau$ masses break axial
$L_{\mu} - L_{\tau}$, the symmetry breaking of the new $U(1)$
is related to electroweak symmetry breaking. While
this is an interesting model, we choose not to examine
it here. The last case has a type of GIM mechanism because the
mixing is maximal. Here decays of the tau such
as $\tau \rightarrow \mu \mu \mu$ are not induced by tree-level
$Z'$ exchange if the states are mass eigenstates also, although
other flavour-changing
decay modes are possible (as we will discuss). In this paper,
it turns out that we will be led to concentrating
on the third possibility, but we will also consider a case
near the end of the paper where the weak and mass eigenstates
are not related in any of the above ways.

We now discuss the model in detail. For second and third generation
leptons we have that
\begin{eqnarray}
& \ell_{2L} \sim (1,2)(-1,2a),\qquad e_{2R} \sim
(1,1)(-2,2a),&\ \nonumber\\
& \ell_{3L} \sim (1,2)(-1,-2a),\qquad e_{3R} \sim
(1,1)(-2,-2a).&\
\label{a_charge}
\end{eqnarray}
It is interesting to note that a number of benefits can be gained
by instituting an exact discrete symmetry under
\begin{equation}
\ell_{2L} \leftrightarrow \ell_{3L},\quad
e_{2R} \leftrightarrow e_{3R},\quad B^{\mu} \leftrightarrow B^{\mu}
\quad {\rm and}\quad Z'^{\mu}
\leftrightarrow -Z'^{\mu},
\end{equation}
where $B^{\mu}$ and $Z'^{\mu}$ are
the gauge fields for U(1)$_Y$ and U(1)$'$, respectively.
The benefits are: (i) It
forces the number of free parameters in the gauge
covariant derivative to be reduced by one, as we explain below.
(ii) If unbroken, it ensures that the
mass eigenstates will be maximally mixed combinations of the weak
eigenstates so that a type of GIM mechanism ensues (as discussed
above). (iii) If unbroken it also forbids $Z-Z'$ mixing to all orders,
which simplifies the phenomenological analysis since the important
experimental constraints on such mixing \cite{Zmixing} are automatically
satisfied.  We will at first be concerned
with the version of the model maintaining
the discrete symmetry as exact. We will then briefly consider
the case where the discrete symmetry is broken by the vacuum.

The gauge covariant derivative for
the electrically neutral gauge bosons is
\begin{equation}
D^{\mu} = \partial^{\mu} + i g_2 I_3 W^{\mu} + i {g_1 \over 2} Y
B^{\mu} + i {g_1 \over 2} Y' Z'^{\mu},
\end{equation}
where $I_3 \equiv \tau_3/2$,
$W^{\mu}$ is the neutral weak-isospin gauge boson and $g_{1,2}$
are the two gauge coupling constants. The coupling constant for U(1)$'$
has been defined to be equal to that for U(1)$_Y$ because the free
parameter $a$ in Eq.~(\ref{a_charge}) can be taken to determine the
relative strengths of these interactions. The parameter that is eliminated
by the discrete symmetry can be identified from an examination of the
kinetic energy Lagrangian for
the U(1) gauge fields. If the discrete symmetry is
ignored then this Lagrangian is given in general by
\begin{equation}
{\cal L}_{\rm KE} = k_1 F^{\mu\nu}F_{\mu\nu} +
k_2 F'^{\mu\nu}F_{\mu\nu} + k_3 F'^{\mu\nu}F'_{\mu\nu}
\end{equation}
where $F$ and $F'$ are the field strength tensors for $B^{\mu}$ and
$Z'^{\mu}$, respectively.
Note that the off-diagonal term is permitted by gauge-invariance because
the symmetries are Abelian \cite{foot-he}.
In order to bring this general kinetic energy
Lagrangian into diagonal and canonically-normalized form we must rewrite
everything in terms of certain linear combinations of $B^{\mu}$ and
$Z'^{\mu}$. If the discrete symmetry is imposed then the mixing term is
absent and so $k_2=0$. In this case the redefinition required is just a
straight rescaling of both $B^{\mu}$ and $Z'^{\mu}$, and the parameters
$k_1$ and $k_3$ can be absorbed by $g_1$ and $a$. If the discrete
symmetry is absent, then the $k_2$ coefficient is an additional free
parameter in the theory (that is, it can {\it not} be absorbed into
$g_1$ and $a$). In the diagonal and conventionally normalised basis for
the gauge fields, the freedom represented by $k_2$ can be incorporated
by the substitution $Y' \to Y' + k Y$ in the covariant derivative where
$k$ is now the arbitrary parameter.

It is convenient to rewrite the gauge covariant derivative in terms of
the photon field $A^{\mu}$ and the standard $Z^{\mu}$ field. The
rewritten covariant derivative is
\begin{equation}
D^{\mu} = \partial^{\mu} + i e Q A^{\mu} + i {e \over {s_W c_W}} (I_3 -
s^2_W Q) Z^{\mu} + i {e \over c_W} {Y' \over 2} Z'^{\mu},
\end{equation}
where $s_W \equiv \sin\theta_W$, $c_W \equiv \cos\theta_W$ and the
weak angle $\theta_W$ is defined
through $\tan\theta_W \equiv g_1/g_2$. The electromagnetic coupling
constant is given as usual by $e \equiv g_2 s_W$ and
electric charge $Q$ is $Q \equiv I_3 + Y/2$. Under the discrete symmetry,
\begin{equation}
A^{\mu} \to A^{\mu},\quad Z^{\mu} \to Z^{\mu}\quad {\rm and}\quad
Z'^{\mu} \to -Z'^{\mu}.
\end{equation}
This means that if the discrete symmetry remains exact after spontaneous
symmetry breaking then there is no $Z-Z'$ mixing to all orders. This is
very important phenomenologically.

The Yukawa coupling Lagrangian is
\begin{equation}
{\cal L}_{\rm Yuk} = \lambda (\overline{\ell}_{2L} e_{2R} \phi_1 +
\overline{\ell}_{3L} e_{3R} \phi_1)\
 + \lambda' (\overline{\ell}_{2L} e_{3R} \phi_2 +
\overline{\ell}_{3L} e_{2R} \phi_3) + {\rm H.c.},
\label{yuk}
\end{equation}
where the Higgs doublet transformation laws are,
\begin{eqnarray}
& \phi_1 \sim (1,2)(1,0),&\ \nonumber\\
& \phi_2 \sim (1,2)(1,4a)\quad
{\rm and}\quad \phi_3 \sim (1,2)(1,-4a).&\
\end{eqnarray}
Under the discrete symmetry $\phi_1 \leftrightarrow \phi_1$ and $\phi_2
\leftrightarrow \phi_3$. (If the discrete symmetry is not invoked then
one need only introduce the equivalent of one of $\phi_2$ and $\phi_3$.)

Let us now look at the phenomenology of the model. First note that
without any analysis there are two potentially very interesting features
of this model due to the hypothesis that the U(1)$'$ breaking is tied
to electroweak breaking. The first is that since the mass of $Z'$
is expected to be less than or equal to the $Z$ boson mass, then the
$Z'$ boson should have some effect on already measurable low energy
observables provided its coupling to leptons is not too weak. Secondly,
if $m_{Z'} < (m_\tau - m_\mu) $ then the two-body decay
$\tau \rightarrow \mu Z'$ can occur. This will provide a striking signature
since the final state muon (in the tauon rest frame) will have a fixed
energy in contrast to the continuum muon energy spectrum from the usual
three-body decay mode. It is known that measurements of the muon energy
spectrum can provide a sensitive probe of the two-body decay mode.
Furthermore, such a signature would easily distinguish this $Z'$ from
that of many other models.

Having foreshadowed what to expect we will now proceed to analyse the
phenomenological implications of the model. We know that $Z-Z'$
mixing is constrained to be small. Our discrete symmetry affords us the
luxury of having this mixing as precisely zero, as discussed above,
provided it is not spontaneously broken. We thus adopt, to begin with,
that range of parameters in the Higgs potential (which we
will display explicitly below) which maintains the discrete symmetry as
exact, while breaking both U(1)$'$ and the electroweak gauge group. The
VEV pattern required is
\begin{equation}
\langle\phi_1\rangle \equiv u_1\ (\neq 0\ {\rm in\ general})\quad
{\rm and}\quad |\langle\phi_2\rangle| = |\langle\phi_3\rangle| =
u_2 \neq 0.
\end{equation}
(We will without loss of generality take the phase of
$\langle\phi_1\rangle$ to be 1, while the phases of
$\langle\phi_2\rangle$ and $\langle\phi_3\rangle$ will be discussed
presently.)
The $Z$ and $Z'$ masses are then given by,
\begin{equation}
m^2_Z = {1 \over 2} (g_1^2 + g_2^2) (u_1^2 + 2 u_2^2)\quad {\rm
and}\quad m^2_{Z'} = 16 a^2 s^2_W (g_1^2 + g_2^2) u_2^2.
\label{mass1}
\end{equation}
As we will soon see, phenomenological bounds force us to consider
the $Z'$ to be much lighter than the $Z$, in contrast to most other $Z'$
models. The reason for this is that in many processes
the parameter $a$ cancels out
between the $Z'$-fermion vertices and the $Z'$ propagator when the
momentum in the propagator can be neglected relative to the $Z'$ mass.
This means that the coupling strength for the $Z'$ in this high mass
limit is completely specified by previously measured quantities. It just
so happens that this coupling strength is {\it too} strong. So, we will
be led to looking at the $m_{Z'} \ll m_Z$ region of parameter space.

Since the discrete symmetry is exact, all mass eigenstate fields have to
be either even or odd under the transformation (this is true for the
neutral gauge bosons discussed above for instance). This allows us to
immediately write down that the mass eigenstate charged leptons are
given by $(e_2 \pm e_3)/\sqrt{2}$. Substituting this into the
$Z'$-lepton interaction Lagrangian we obtain that
\begin{equation}
{\cal L}^{\ell}_{\rm int} = {{ea} \over c_W}
(\overline{\mu} \gamma^{\mu} \tau
+ \overline{\tau} \gamma^{\mu} \mu) Z'_{\mu}.
\label{L1}
\end{equation}
Therefore we see that although the $Z'$ boson mediates flavour-changing
neutral currents, these processes are always {\it purely off-diagonal}.
(Diagonal terms are forbidden because $Z'$ is odd.) By defining muon and
tau neutrinos as those fields that are produced with muons and tau
leptons, respectively, in charged current weak interactions we also see
that
\begin{equation}
{\cal L}^{\nu}_{\rm int} = {{ea} \over 2 c_W}
(\overline{\nu}_{\mu} \gamma^{\mu} (1-\gamma_5)\nu_{\tau}
+ \overline{\nu}_{\tau} \gamma^{\mu} (1-\gamma_5)\nu_{\mu}) Z'_{\mu}.
\label{L2}
\end{equation}
These Lagrangians will allow us to easily identify the interesting
phenomenological constraints on $Z'$-lepton interactions.

The most convenient way to write the Higgs potential down is
\begin{eqnarray}
V & = &\lambda_1 ( \phi^{\dagger}_1 \phi_1 - u_1^2 )^2 + \lambda_2 (
\phi_2^{\dagger} \phi_2 + \phi_3^{\dagger} \phi_3 - 2 u_2^2 )^2\nonumber
\\
& + & \lambda_3 ( \phi_2^{\dagger} \phi_2 - \phi_3^{\dagger} \phi_3 )^2
+ \lambda_4 ( \phi^{\dagger}_1 \phi_1 + \phi_2^{\dagger} \phi_2 +
\phi_3^{\dagger} \phi_3 - u_1^2 - 2 u_2^2 )^2\nonumber\\
& + & \lambda_5 [ (\phi_2^{\dagger} \phi_2)(\phi_3^{\dagger} \phi_3) -
(\phi_2^{\dagger} \phi_3)(\phi_3^{\dagger} \phi_2) ]\nonumber\\
& + & \lambda_6 [ \phi^{\dagger}_1 \phi_1 ( \phi_2^{\dagger} \phi_2 +
\phi_3^{\dagger} \phi_3 ) - (\phi_1^{\dagger} \phi_2)(\phi_2^{\dagger}
\phi_1) - (\phi_1^{\dagger} \phi_3)(\phi_3^{\dagger} \phi_1) ]\nonumber\\
& + & \lambda_7 [ \phi^{\dagger}_1\phi_1 ( \phi_2^{\dagger} \phi_2 +
\phi_3^{\dagger} \phi_3 ) - (\phi_1^{\dagger} \phi_2)(\phi_1^{\dagger}
\phi_3) - (\phi_2^{\dagger} \phi_1)(\phi_3^{\dagger} \phi_1)].
\label{pot1}
\end{eqnarray}
The parameters $\lambda_{1-7}$ must be real from hermiticity
(we have also redefined to zero a phase that can a priori appear
in front of the last two terms within the $\lambda_7$ term). The
symmetry breaking pattern we require is obtained by choosing
$\lambda_{1-7} > 0$. (Other
symmetry breaking patterns can of course
be induced in other regions of parameter space.)
In the $\lambda_{1-7} > 0$ region of parameter space,
the Higgs potential is
the sum of positive-definite terms. The $\lambda_{1-4}$ terms
are obviously positive-definite, while a little algebra shows that
\begin{eqnarray}
(\phi_2^{\dagger} \phi_2)(\phi_3^{\dagger} \phi_3) -
(\phi_2^{\dagger} \phi_3)(\phi_3^{\dagger} \phi_2) & = &
| \phi^+_3 \phi^0_2 - \phi^+_2 \phi^0_3 |^2,\nonumber\\
(\phi_1^{\dagger} \phi_1)(\phi_2^{\dagger} \phi_2) -
(\phi_1^{\dagger} \phi_2)(\phi_2^{\dagger} \phi_1) & = &
| \phi^+_2 \phi^0_1 - \phi^+_1 \phi^0_2 |^2,\nonumber\\
(\phi_1^{\dagger} \phi_1)(\phi_3^{\dagger} \phi_3) -
(\phi_1^{\dagger} \phi_3)(\phi_3^{\dagger} \phi_1) & = &
| \phi^+_3 \phi^0_1 - \phi^+_1 \phi^0_3 |^2,
\label{charged_vevs}
\end{eqnarray}
and
\begin{eqnarray}
\phi^{\dagger}_1\phi_1 ( \phi_2^{\dagger} \phi_2 +
\phi_3^{\dagger} \phi_3 ) & - & (\phi_1^{\dagger} \phi_2)(\phi_1^{\dagger}
\phi_3) - (\phi_2^{\dagger} \phi_1)(\phi_3^{\dagger}\phi_1) \nonumber\\
& = & | \phi^+_1 \phi^-_2 - \phi^-_1 \phi^+_3 |^2 +
| \phi^+_1 \phi^{0*}_2 - \phi^{0*}_1 \phi^+_3 |^2\nonumber\\
& + & | \phi^+_1 \phi^{0*}_3 - \phi^{0*}_1 \phi^+_2 |^2
+ | \phi^0_1 \phi^{0*}_2 - \phi^{0*}_1 \phi^0_3 |^2.
\label{lambda_4}
\end{eqnarray}
Since the Higgs potential is written as the sum of positive-definite
terms, we know we have a minimum if a VEV pattern renders each term
separately zero. The first four terms show that $|\langle\phi_1\rangle|
= u_1$, and $|\langle\phi_2\rangle| = |\langle\phi_3\rangle| = u_2$. The
$\lambda_3$ term is responsible for forcing the last two VEVs to be
exactly equal. The $\lambda_{5,6}$ terms force the charged Higgs bosons
to have zero VEVs. To see this, first perform an SU(2)$_L$ gauge
transformation to define $\langle\phi_1^{\pm}\rangle = 0$. Then
$\langle\phi_1^0\rangle = u_1 [\neq 0\ {\rm by\ parameter\ choice}$, and
it can be made positive and real by the same SU(2)$_L$ transformation].
Minimization of the terms in
Eq.~(\ref{charged_vevs}) forces $\langle\phi_{2,3}^{\pm}\rangle = 0$.
The first three terms on the right-hand side of Eq.~(\ref{lambda_4})
are then also zero, while the fourth
term tells us that $\langle\phi^0_1\rangle
\langle\phi^{0*}_2\rangle = \langle\phi^{0*}_1\rangle
\langle\phi^0_3\rangle$. Given that the phase angle for $\phi_1$ has
been set to 1, this implies that the phase angles for $\phi_2$ and
$\phi_3$ are equal and opposite. However, these phase angles can be
removed by a U(1)$'$ transformation and are thus unphysical and will
henceforth be set to zero.

Consider the shifted neutral Higgs fields defined through
\begin{equation}
\phi^0_1 \equiv u_1 + {{H_1 + i \eta_1} \over \sqrt{2}},
\quad {\rm and}\quad
\phi^0_{2,3} \equiv u_2 + {{H_{2,3} + i \eta_{2,3}} \over \sqrt{2}},
\end{equation}
where the $H$'s are CP-even real Higgs bosons and the $\eta$'s are
CP-odd real Higgs bosons. It is convenient to discuss the $\phi_{2,3}$
fields in the discrete symmetry eigenstate basis given by
\begin{equation}
H_{\pm} \equiv {{H_2 \pm H_3} \over \sqrt{2}}\quad {\rm and}\quad
\eta_{\pm} \equiv {{\eta_2 \pm \eta_3} \over \sqrt{2}},
\end{equation}
where the subscripts $+$ and $-$ denote even and odd fields under the
discrete symmetry, respectively.

All three $H$-fields are physical. The odd combination $H_{-}$ does not
mix with the even fields $H_1$ and $H_{+}$. The mass of the former is
\begin{equation}
m^2_{H_{-}} = 8 \lambda_3 u_2^2 + 2 \lambda_7 u_1^2,
\end{equation}
while the mass matrix for the latter two is
\begin{equation}
m^2(H_1, H_{+}) = \left(
\begin{array}{cc}
4(\lambda_1+\lambda_4)u_1^2\ \ & 4\sqrt{2}\lambda_4u_1u_2 \\
4\sqrt{2}\lambda_4u_1u_2\ \ & 8(\lambda_2+\lambda_4)u_2^2
\end{array} \right).
\end{equation}
There is one physical $\eta$-field given by
\begin{equation}
\eta_{\rm Phys} =
{{\sqrt{2}u_2\eta_1 - u_1\eta_{+}} \over \sqrt{u_1^2+2u_2^2}}
\end{equation}
with mass
\begin{equation}
m^2_{\eta_{\rm Phys}} = 2\lambda_7 (u_1^2+2u_2^2).
\end{equation}
There are two physical charged Higgs bosons, given by
\begin{equation}
h^+ \equiv {{\sqrt{2}u_2\phi_1^+ - u_1 h^+_+} \over
\sqrt{u_1^2+2u_2^2}}\quad {\rm and}\quad h^+_- \equiv {{\phi_2^+ -
\phi_3^+} \over \sqrt{2}},
\end{equation}
where $h^+_+ \equiv (\phi_2^+ + \phi_3^+)/\sqrt{2}$. Their masses are
\begin{equation}
m^2_{h^+} = (\lambda_6 + \lambda_7)(u_1^2 + 2u_2^2)\quad {\rm and}\quad
m^2_{h_-^+} = (\lambda_6 + \lambda_7)u_1^2 + 2 \lambda_5 u_2^2.
\end{equation}
Note that the odd combination $h_-^+$ does not mix with the even
combination $h^+$ because of the exact discrete symmetry.

The Yukawa coupling Lagrangian for the $H$-fields is
\begin{eqnarray}
{\cal L}^H_{\rm Yuk} & = & \sum_{f} {m_f \over {\sqrt{2}u_1}}
\overline{f} f H_1 +
{{m_{\tau}+m_{\mu}} \over {2\sqrt{2}u_1}}
(\overline{\tau} \tau + \overline{\mu} \mu) H_1\nonumber\\
& + & {{m_{\tau}-m_{\mu}} \over {2\sqrt{2}u_2}} \left[
(\overline{\tau} \tau -
\overline{\mu} \mu) H_+ + (\overline{\mu} \tau -
\overline{\tau} \mu) H_- \right],
\end{eqnarray}
where $f = u,d,c,s,t,b,e$ and $m_f$ is the corresponding mass.
The Lagrangian for the physical $\eta$-field is
\begin{eqnarray}
{\cal L}^{\eta}_{\rm Yuk} & = & \sum_{f}
{{iu_2 m_f} \over {u_1\sqrt{u_1^2+2u_2^2}}} \overline{f} \gamma_5
f \eta_{\rm Phys}\nonumber\\
& + & {i \over 2\sqrt{u_1^2+2u_2^2}}
\left[ {u_2 \over u_1} (m_{\tau}+m_{\mu})
- {u_1 \over 2u_2} (m_{\tau}-m_{\mu}) \right]\  \overline{\tau} \gamma_5
\tau\  \eta_{\rm Phys}\nonumber\\
& + & {i \over 2\sqrt{u_1^2+2u_2^2}}
\left[ {u_2 \over u_1} (m_{\tau}+m_{\mu})
+ {u_1 \over 2u_2} (m_{\tau}-m_{\mu}) \right]\  \overline{\mu} \gamma_5
\mu\  \eta_{\rm Phys}.
\end{eqnarray}
It is interesting to note that the mass eigenstate CP-even Higgs bosons
that are superpositions of $H_1$ and $H_+$ have flavour-diagonal
interactions, as does the physical CP-odd field $\eta_{\rm Phys}$. The
discrete-symmetry-odd mass eigenstate $H_-$ is flavour-changing in the
$\mu$-$\tau$ sector, but it is completely off-diagonal just like the
$Z'$ (and for the same reason of course).

The Lagrangian for fermion coupling to the charged Higgs bosons is
\begin{eqnarray}
{\cal L}^+_{\rm Yuk} & = & {{\sqrt{2}u_2} \over {u_1\sqrt{u_1^2+2u_2^2}}}
\overline{U}_L M_d D_R h^+
+ {{\sqrt{2} u_2 m_e} \over {u_1\sqrt{u_1^2+2u_2^2}}} \overline{\nu}_{eL}
e_R h^+\nonumber\\
& + & {1 \over {\sqrt{2}\sqrt{u_1^2+2u_2^2}}} \left[ {u_2 \over u_1}
(m_{\tau}+m_{\mu}) - {u_1 \over 2u_2} (m_{\tau}-m_{\mu}) \right]\
\overline{\nu}_{\tau L} \tau_R\  h^+ \nonumber\\
& + & {1 \over {\sqrt{2}\sqrt{u_1^2+2u_2^2}}}\left[ {u_2 \over u_1}
(m_{\tau}+m_{\mu}) + {u_1 \over 2u_2} (m_{\tau}-m_{\mu}) \right]\
\overline{\nu}_{\mu L} \mu_R\ h^+ \nonumber\\
& + & {{m_{\tau}-m_{\mu}} \over {2\sqrt{2}u_2}} (\overline{\nu}_{\mu L}
\tau_R - \overline{\nu}_{\tau L} \mu_R) h_-^+ + {\rm H.c.}
\end{eqnarray}
where $U \equiv (u,c,t)$, $D^T \equiv (d,s,b)$ and $M_d$ is the
undiagonalised down-quark mass matrix.

We now have to identify those processes involving second and third
generation leptons which provide significant phenomenological
constraints. (Since $Z$-$Z'$ mixing is absent to all orders, the number
of relevant processes is greatly reduced.) There are essentially three
important constraints: the anomalous magnetic moment of the muon, $a_{\mu}$,
the gauge boson masses (i.e., we have to ensure that the values of $u_1$
and $u_2$ reproduce the measured values for $m_W$ and $m_Z$ and hence
$m_{Z'}$ cannot be arbitrarily large)
and the $Z'$ contribution to $\tau$ decay \cite{pdg}.

The principal contribution to $a_{\mu}$ is depicted in Fig.~1. Note
that the discrete-symmetry-odd field $Z'$ is featured here, because it
couples muons to tau leptons. There are similar graphs involving the
neutral and charged Higgs bosons which also contribute to $a_{\mu}$,
but they turn out to be much smaller since they are suppressed by the
factor $\left(m_\mu / m_H \right)^2$ for $m_\mu < m_H$ where $m_H$ is
the mass of the generic Higgs field. For the analysis that follows
we will assume that all the Higgs bosons in the model are heavier than
${\cal O}(40)$ GeV so that their contributions to $a_{\mu}$ and the
decay width of the standard $Z$ boson are negligible.

The $Z'$ contribution to $a_{\mu}$ is given by,
\begin{equation}
\Delta a^{Z'}_{\mu} = {\alpha_{\rm em} \over 2\pi} {|a|^2 \over
c^2_W} \left\{ \gamma + 2(\beta - {2B\over C}\gamma)
+ 2M\ln\left({m_\tau \over m_{Z'}}\right) + \delta \right\},
\label{g-2Z'}
\end{equation}
where
\begin{equation}
\delta = \left\{
\begin{array}{ll}
{{NC-MB} \over \sqrt{B^2-AC}} \ln\left|
{{A+B+\sqrt{B^2-AC}} \over {A+B-\sqrt{B^2-AC}}}\right|
& \mbox{if $B^2>AC$;} \\
2{{NC-MB} \over \sqrt{AC-B^2}} \tan^{-1}\left[{\sqrt{AC-B^2}
\over {A+B}}\right] & \mbox{if $B^2<AC$.}
\end{array} \right.
\end{equation}
In these equations, $\alpha_{\rm em}$ is the fine-structure constant,
\begin{eqnarray}
&\alpha \equiv 2({m_{\tau} - m_{\mu}})/m_{\mu}, &\ \nonumber\\
&\beta \equiv 3 - 2(m_{\tau}/ m_\mu) +
{1\over 2 }\left((m_{\tau}/ m_{\mu})+1\right)
(m_{\tau}-m_{\mu})^2/ m_{Z'}^2,&\ \nonumber\\
&\gamma \equiv -1 - {1\over 2}(m_{\tau}-m_{\mu})^2/m^2_{Z'},&\ \nonumber\\
&A \equiv m^2_{Z'},\quad B \equiv (m^2_{\tau}-m^2_{\mu}-m^2_{Z'} )/2,
\quad C \equiv m^2_{\mu},&\ \nonumber\\
&M \equiv \alpha - {2B\over C}(\beta-{2B\over C}\gamma)\quad {\rm and}
\quad N \equiv -{A\over C}(\beta-{2B\over C}\gamma).&\
\end{eqnarray}
This expression demonstrates that a large $m_{Z'}$ is phenomenologically
disallowed. In the $m_{Z'} \gg m_{\tau}$ limit, Eq.~(\ref{g-2Z'})
reduces to the simple result that
\begin{equation}
\Delta a^{Z'}_{\mu} \simeq {\alpha_{\rm em} \over 2\pi} {|a|^2 \over
c^2_W}{{2m_{\mu}m_{\tau}} \over m^2_{Z'}} = {m_{\mu}m_{\tau} \over
64\pi^2u^2_2}
\end{equation}
which is independent of $|a|$ and at best about an order of magnitude
too large given that $u_2$ is constrained by the weak scale. So, we
will be interested in $Z'$ masses of about a few GeV or less.
In any case Eq.~(\ref{g-2Z'}) can be evaluated numerically.
Figures 2(a) and (b) show the allowed region of $(m_{Z'}, |a|)$
parameter space, given the experimental constraint \cite{pdg}
$|\Delta a_{\mu}| <
10^{-8}$, i.e., given by the region below the dashed curve.

We next consider the constraint coming from the gauge boson masses.
By using Eqs.~(\ref{mass1}) and $m_W^2 = g_2^2 (u_1^2 + 2u_2^2)/2$,
one finds \cite{footnote1}
\begin{equation}
|a| > {1\over 4\tan\theta_W}{m_{Z'}\over m_W} \simeq \left({m_{Z'}
\over 175.33 \ \hbox{GeV}}\right).
\label{bd}
\end{equation}
The region allowed by this constraint is the area {\it above} the solid
curve shown in Figs. 2(a) and (b). In other words, for a given
value of $m_{Z'}$ there exists a minimum value for $|a|$.
When the above two constraints are combined there is a small overlap
region remaining. This overlap region where the two constraints are
satisfied (roughly $m_{Z'} < 2.5$ GeV) divides into regimes; namely,
$m_{Z'} > (m_{\tau} - m_{\mu})$ and $m_{Z'} < (m_{\tau} - m_{\mu})$.
In the former, the interesting decay mode $\tau \to \mu Z'$ is not
allowed kinematically, whereas in the latter it is. Note that this
result makes numerically precise the qualitative observation made
earlier that the $Z'$ boson cannot be arbitrarily heavy.

Let us first consider the case where $\tau \to \mu Z'$ is not allowed.
Although this dramatic two-body decay does not occur, the off-shell $Z'$
contributes to the family lepton-number preserving
three-body decay $\tau^- \to \mu^- \overline{\nu}_{\mu}
\nu_{\tau}$ and the family lepton-number violating decay
$\tau^- \to \mu^- \nu_{\mu} \overline{\nu}_{\tau}$. We have to check
whether or not constraints from the observation of the standard decay
mode $\tau^- \to \mu^- \overline{\nu}_{\mu} \nu_{\tau}$ close the
$(m_{\tau}-m_{\mu}) < m_{Z'} < 2.5$ GeV window. For this mode, the $Z'$
contribution coherently adds with the standard $W$-boson contribution
yielding
\begin{eqnarray}
R & \equiv & {\Gamma(\tau^- \to \mu^- \overline{\nu}_{\mu} \nu_{\tau}) \over
\Gamma(\tau^- \to \mu^- \overline{\nu}_{\mu} \nu_{\tau})_{\rm SM}}
\nonumber \\
& = & 1 - \xi \left[ 2k(k+1) - {5 \over 6}
- k^2(2k+3)\ln \left|{1+k \over k} \right| \ \right]\nonumber\\
& + & { 1\over 4} \xi^2 \left[ 2(2k+1) + k{2k+3 \over k+1} - 6k(k+1) \ln
\left|{1+k \over k}\right| \ \right],
\label{3body}
\end{eqnarray}
where
\begin{equation}
k \equiv {m^2_{Z'} \over m^2_{\tau}} - 1\quad {\rm and}\quad \xi \equiv
{\sqrt{2} \over G_Fm^2_{\tau}} {4\pi\alpha_{\rm em} \over c^2_W} |a|^2.
\end{equation}
Note that the contribution from the finite width of $Z'$ has been neglected
in this calculation. This contribution is expected to have its most
significant effect near the $Z'$ threshold. However, in this case,
the $Z'$ width is given by
\begin{equation}
\Gamma_{Z'} \simeq
\Gamma(Z' \rightarrow  \overline{\nu}_{\mu} \nu_{\tau}
+  \nu_{\mu} \overline{\nu}_{\tau})
= {\alpha_{\rm em} \over 3 c_W^2}|a|^2 m_{Z'}
\end{equation}
where it is supressed by a factor of $|a|^2$ so that the zero width
approximation should not be a bad one.
The largest contribution comes from the interference term between the $W$
and $Z'$ bosons. (The non-standard decay $\tau^- \to \mu^- \nu_{\mu}
\overline{\nu}_{\tau}$ mode will always provide less stringent
constraints than the $Z'$ contribution to the standard decay because the
decay rate is given by the direct-$Z'$ process only and is thus proportional
to $|a|^4$.) The experimental constraint \cite{pdg}
\begin{equation}
|R-1| < 0.04
\end{equation}
in fact closes the $(m_{\tau}-m_{\mu}) < m_{Z'} < 2.5$ GeV window.
This is shown in Figs. 2(a) and (b) where the region below the
dot-dashed curve is the one allowed by the three-body decay constraint.

So, we are left to consider the kinematic region which permits the
two-body decay mode $\tau \to \mu Z'$. Firstly, notice that the
three-body decay constraint allows for windows in the $m_{Z'} < 0.2$
GeV and $0.8 < m_{Z'} < 1.0$ GeV regions. There is also a minute region
at $m_{Z'} \sim 1.2$ GeV. The second window is caused by the vanishing
of the term proportional to $\xi$ in Eq.~(\ref{3body}) for values
of $m_{Z'}$ in this region, while the third window is due to the cancellation
between the $\xi$ and $\xi^2$ terms in Eq.~(\ref{3body}). (This
cancellation is possible for large enough values of $|a|$ because
the $\xi^2$ term becomes as important as the $\xi$ term.)
We now check to see what effect the two-body decay mode has.
The Mark III and ARGUS collaborations \cite{exp} have set limits on two-body
decay modes for $\tau$. These experimental groups specifically analysed
the process $\tau \to \mu +\ \hbox{Goldstone Boson}$ and found that the
ratio
\begin{equation}
{\Gamma(\tau \rightarrow \mu Z') \over
\Gamma(\tau \rightarrow \mu\overline{\nu_\mu} \nu_\tau)}
< 0.033, \quad \hbox{for} \ m_{Z'} \leq 0.1 \ \hbox{GeV},
\label{2bb}
\end{equation}
where the Goldstone boson has been replaced by $Z'$.
(Without going into a detailed reanalysis of the experiment,
we expect the above experimental bound to be approximately valid
for our case where the final state boson has spin-1.)
This bound rises up to $0.071$ for $m_{Z'} = 0.5$ GeV. For the
exact discrete symmetry case this ratio is given by
\begin{equation}
{\Gamma(\tau \rightarrow \mu Z') \over
\Gamma(\tau \rightarrow \mu\overline{\nu_\mu} \nu_\tau)}
= {96\over\sqrt{2}}\pi^2 \tan^2\theta_W
{m_W^2\over G_F m_{\tau}^4} |a|^2 f,
\label{2body}
\end{equation}
where
\begin{equation}
f=\left\{1+{(m_\mu^2-2m_{Z'}^2)\over m_{\tau}^2}-6{m_\mu\over m_\tau}
+{(m_\mu^2-2m_{Z'}^2)^2\over m_{\tau}^2 m_{Z'}^2}\right\} PS
\end{equation}
and
\begin{equation}
PS=\sqrt{1-{(m_\mu+m_{Z'})^2\over m_{\tau}^2}}
\sqrt{1-{(m_\mu-m_{Z'})^2\over m_{\tau}^2}}.
\end{equation}
$G_F$ is the Fermi constant and $m_W$ is the mass of the $W$ boson.
By using Eqs.~(\ref{2bb}) and (\ref{2body}), the region of
$(m_{Z'}, |a|)$ parameter space allowed by the two-body decay
can be constructed. This is given by the region below the dotted
curve in Fig. 2(b). From this one can see that the parameter space
for $m_{Z'} < 0.5$ GeV is ruled out ( and hence the window of
$m_{Z'} < 0.2$ GeV allowed by the three-body constraint).

So, in summary, when all the constraints have been combined,
much of the parameter space is ruled out. The remaining
allowed regions are for $0.8 < m_{Z'} < 1.0$ GeV ($|a|$ varies
between about $0.004$ and $0.007$) and $m_{Z'}$
around $1.2$ GeV. [One might naively think that there ought to
be another allowed region for sufficiently small $|a|$, and hence for a
sufficiently light $Z'$ boson, since the $Z'$ decouples as $|a| \to 0$.
However, as $|a| \to 0$ the local U(1)$'$ gauge symmetry tends toward
becoming merely a global symmetry, and the longitudinal component of the
$Z'$ turns into its associated Goldstone boson. The 2-body decay
process considered above then has this Goldstone boson in the final
state rather than the $Z'$. This can be seen explicitly from the fact
that the right-hand side Eqn.~(39) does not go to zero as $|a|$ goes to
zero.] It should be noted that we have
taken the two-body constraint at face value, i.e., it applies for
values of $m_{Z'}$ up to $0.5$ GeV. This is the value quoted by the
ARGUS collaboration in Ref.~\cite{exp}. Actually, the ARGUS experiment
is supposed to be able to search for the two-body decay mode for values
of $m_{Z'}$ up to about $1.53$ GeV, given the experimental cuts and
efficiencies. If the current trend of the two-body constraint continues
beyond $0.5$ GeV (the precise bound will obviously vary with the mass of
$Z'$ and becomes several orders of magnitude less severe near threshold)
then the remaining allowed windows will be closed and the model will be
ruled out.

We now discuss what happens when the discrete symmetry is
spontaneously broken by the vacuum. The calculations of the
constraints are similar to those in the unbroken discrete
symmetry case (see the appendix for further details). For the gauge
boson mass constraint, the calculation uses the mass relations of
Eqs.~(\ref{gaugemass})--(\ref{mix}) in the appendix. The calculation
of the anomalous magnetic moment \cite{footnote2},
two-body decay and
three-body decay constraints can be carried over from the exact
discrete symmetry case using the substitution
$|a| \rightarrow |a| \cos\phi\sin 2\theta_L$, where $\phi$ and
$\theta_L$ are the gauge boson and $\mu$-$\tau$ mixing angles respectively.
(Since the gauge boson mixing is required to be small, we have set
$\phi \simeq 0$.)
The results are given in Figs. 3(a) and (b) which shows that the
broken discrete symmetry case is qualitatively similar to the exact
discrete symmetry case. So the conclusions made for the exact discrete
symmetry case essentially also hold for this case. The constraints were
expected to be less stringent, which they are, but not enough to
dramatically change anything significantly. For example, there is
still no allowed parameter space for $m_{Z'} > (m_\tau - m_\mu)$
since the allowed regions from the gauge boson mass and three-body
constraints never overlap for $m_{Z'} > 1.5$ GeV. The reason for
this is due to the fact that $\sin 2\theta_L$ cannot be made
arbitrarily small [this is a consequence of the discrete symmetry
of the Yukawa Lagrangian of Eq.~(\ref{yuk})]. It turns out that
$\sin 2\theta_L$ cannot be smaller than about 0.46 [see Eq.~(\ref{sinmin})
in the appendix]. Therefore, one obvious way to ease the constraints
is to abandon the discrete symmetry altogether so that the $\mu-\tau$
mass mixing remains unconstrained.

In conclusion then, the model for both the exact and spontaneously
broken discrete symmetry cases is ruled out if $m_{Z'} \leq 0.5$ GeV
or $m_{Z'} > (m_\tau -m_\mu)$. For $0.5\ \hbox{GeV}\ < m_{Z'} \leq
(m_\tau -m_\mu)$ there exists windows of allowed parameter space.
However, if the trend of the two-body decay constraint continues in
this region, then these windows will certainly be closed.
{\it This rather stringent bound from the two-body decay is,
nevertheless, very interesting, because it means that the decay
$\tau \to \mu Z'$ is by far the best way to test our low-mass $Z'$ model}.
One way to view the significance of our model is therefore the following:
One should as a matter of phenomenological generality be interested in
the possibility that $\tau$ might have a rare decay mode into $\mu$ plus
a spin-1 boson, just as one is in general interested in two-body final
states where the boson has spin-0. Our model provides a simple model where
this phenomenological possibility is realised. The interesting thing is
that the $\tau \to \mu Z'$ decay is essentially the only important piece
of new low energy physics that the model predicts, provided that the
Higgs bosons are heavier than a few tens of a GeV.

\vskip 1.5cm
\leftline{\bf Acknowledgements}
RF and RRV would like to thank Donald Shaw for some discussions.

\vskip 1.5cm
\leftline{\bf Appendix: Spontaneously broken discrete symmetry}
\vskip 0.5cm
\noindent
In this appendix some details concerning the model with
spontaneously broken discrete symmetry are given. The results
given in the following are the ones used to calculate the
constraints discussed in the text.

\vskip 0.5cm
\leftline{\bf The gauge boson and fermion sector:}
\vskip 0.5cm

When the Higgs doublets develop nonzero VEVs, i.e.,
$|\langle \phi_i \rangle | = u_i$ for $i=1, 2, 3$,
the electroweak and U(1)$'$ symmetries are broken.
This results in a neutral gauge boson mass(-squared)
matrix given by
\begin{equation}
{1\over 2}{e^2\over c_W^2 s_W^2} \left[
\begin{array}{cc}
(u_1^2+u_2^2+u_3^2) & -4a s_W(u_2^2-u_3^2) \\
 -4a s_W(u_2^2-u_3^2) & 16 a^2 s_W^2(u_2^2+u_3^2)
\end{array} \right]
\label{gaugemass}
\end{equation}
in the $(Z, Z')$ basis. In terms of mass eigenstates $(Z_1, Z_2)$
\begin{eqnarray}
Z & = & \cos\phi Z_1 + \sin\phi Z_2, \nonumber\\
Z' & = & -\sin\phi Z_1 + \cos\phi Z_2,
\label{estates}
\end{eqnarray}
where $\phi$ is the $Z-Z'$ mixing angle and is given by
\begin{equation}
\tan 2\phi = {8a s_W(u_2^2-u_3^2)\over
(u_1^2+u_2^2+u_3^2)-16 a^2 s_W^2(u_2^2+u_3^2)}
\label{mix}
\end{equation}
In the exact discrete symmetry limit ($u_2=u_3$), the above
reduces to that given in Eq.~(\ref{mass1}).
The mass of the charged $W^\pm$ boson is
$m_W^2 = {1\over 2} g_2^2 (u_1^2+u_2^2+u_3^2)$.

 From the Yukawa Lagrangian of Eq.~(\ref{yuk}), the $\mu-\tau$
mass matrix can by written as follows:
\begin{equation}
{\cal L}_{\rm mass} = \overline{L_L} {\cal M} L_R + {\rm H.c.,}
\end{equation}
where
\begin{equation}
L_{L,R} = \left(\mu_{L,R}, \tau_{L,R}\right)^T,
\end{equation}
\begin{equation}
{\cal M} = \left( \begin{array}{cc}
m_1 & m_2 \\
m_3 & m_1
\end{array} \right)
\end{equation}
and $m_1 = \lambda u_1$, $m_2 = \lambda' u_2$ and $m_3 = \lambda' u_3$.
The matrix $\cal M$ can be diagonalized by a bi-unitary transformation
so that
\begin{eqnarray}
{\cal D} & = & {\rm Diag}(m_\mu, m_\tau)
= U_L {\cal M} U_R^\dagger, \\
L'_{L,R} & = & U_{L,R} L_{L,R},
\end{eqnarray}
where the $L'_{L,R}$ denotes the mass eigenstates.
$U_{L,R}$ can be parametrized as
\begin{equation}
 U_{L,R} = \left( \begin{array}{cc}
\cos\theta_{L,R} & \sin\theta_{L,R} \\
-\sin\theta_{L,R} & \cos\theta_{L,R} \end{array} \right)
 \left( \begin{array}{cc}
e^{-i{\delta\over 2}} & 0 \\
0 & e^{i{\delta\over 2}} \end{array} \right),
\end{equation}
where
\begin{equation}
\tan 2\theta_L = {2(m_1m_3^* + m_1^* m_2)\over
|m_3|^2 - |m_2|^2} = - \tan 2\theta_R,
\end{equation}
\begin{equation}
\theta_R = (2n+1){\pi\over 2} - \theta_L,
\end{equation}
where $n$ is an integer. In the exact discrete symmetry limit
$\theta_L = \theta_R = {\pi \over 4}$ and $\delta = 0$.
By using the above relations one can rewrite $\theta_L$
in terms of the VEVs and the $\mu$ and $\tau$ masses such that
\begin{equation}
\cos 2\theta_L = {(m_\tau^2 - m_\mu^2) \over (u_3^2-u_2^2)}
{(u_3^2-u_2^2)^2 - 4u_2^2u_3^2\sin^22\delta \over
\left[u_2^2+u_3^2+2u_2u_3\cos2\delta\right]
\left[m_\tau^2+m_\mu^2+2m_\tau m_\mu\cos\Delta \right]}
\end{equation}
where
\begin{equation}
\sin\Delta = {u_2 u_3\over u_2^2 -u_3^2}
{m_\tau^2-m_\mu^2 \over m_\tau m_\mu} \sin 2\delta.
\end{equation}
Furthermore, one can show that
\begin{equation}
|\cos2\theta_L| \leq {m_\tau - m_\mu \over m_\tau + m_\mu}
\quad \Rightarrow \quad
|\sin2\theta_L| \geq 0.46
\label{sinmin}
\end{equation}

Using the foregoing results, the neutral current gauge
interactions can be written as
\begin{eqnarray}
{\cal L}^Z_{\rm int} & = &
-{e\over c_W s_W} \overline{f} \gamma_\mu Z^\mu
(I_3 - s_W^2 Q )P_{L,R} f, \\
{\cal L}^{Z'}_{\rm int} & = &
-{e\over c_W }a \overline{L} \gamma_\mu {Z'}^\mu
\left(\begin{array}{cc}
-\gamma_5 \cos 2\theta_L & -\sin 2\theta_L \\
-\sin 2\theta_L & \gamma_5 \cos 2\theta_L \end{array}\right) L
\nonumber \\
& - & {e\over c_W }a \overline{N} \gamma_\mu {Z'}^\mu
\left(\begin{array}{cc}
\cos 2\theta_L & -\sin 2\theta_L \\
-\sin 2\theta_L & -\cos 2\theta_L \end{array}\right)P_L N,
\end{eqnarray}
where
$P_{L,R}={1\over 2}(1\pm \gamma_5)$,
$ L = \left(\mu, \tau\right)^T$,
$ N = \left(\nu_\mu, \nu_\tau\right)^T$
and $f = L \hbox{ or } N$. The $Z$ and $Z'$ fields can
be written in terms of their mass eigenstates by using
Eq.~(\ref{estates}). Note that these interactions reduce
to the simple form of Eqs.~(\ref{L1}) and (\ref{L2}) in
the exact discrete symmetry limit.

\vskip 0.5cm
\leftline{\bf The Higgs boson sector:}
\vskip 0.5cm

The Higgs potential is given by
\begin{eqnarray}
V(\phi_1,\phi_2,\phi_3) & = &
-\mu_1^2(\phi_1^\dagger\phi_1)
-\mu_2^2(\phi_2^\dagger\phi_2+\phi_3^\dagger\phi_3) \nonumber\\
& + & k_1(\phi_1^\dagger\phi_1)^2
+k_2\left[(\phi_2^\dagger\phi_2)^2+(\phi_3^\dagger\phi_3)^2\right] \nonumber\\
& + & k_{12}(\phi_1^\dagger\phi_1)
\left[(\phi_2^\dagger\phi_2)+(\phi_3^\dagger\phi_3)\right] \nonumber\\
& + & k'_{12}\left[(\phi_1^\dagger\phi_2)(\phi_2^\dagger\phi_1)
+(\phi_1^\dagger\phi_3)(\phi_3^\dagger\phi_1)\right] \nonumber\\
& + & k_{23}(\phi_2^\dagger\phi_2)(\phi_3^\dagger\phi_3)
+ k'_{23}(\phi_2^\dagger\phi_3)(\phi_3^\dagger\phi_2) \nonumber\\
& + & k_4 Re(\phi_1^\dagger\phi_2)(\phi_1^\dagger\phi_3)
+ k'_4 Im(\phi_1^\dagger\phi_2)(\phi_1^\dagger\phi_3).
\label{pot2}
\end{eqnarray}
The minimization conditions are given by
\begin{eqnarray}
0 & = & -\mu_1^2 + 2k_1u_1^2 +2\tilde{k}_{12}\left(u_2^2+u_3^2
\right)+k_4u_2u_3, \nonumber \\
0 & = & 2\left(-\mu_2^2 + 2k_2u_2^2 +2\tilde{k}_{12}u_1^2
+2\tilde{k}_{23}u_3^2 \right)u_2 +k_4u_1^2u_3, \nonumber \\
0 & = & 2\left(-\mu_2^2 + 2k_2u_3^2 +2\tilde{k}_{12}u_1^2
+2\tilde{k}_{23}u_2^2 \right)u_3 +k_4u_1^2u_2 ,
\end{eqnarray}
together with $k'_4 = 0$, where
\begin{equation}
\tilde{k}_{12} \equiv {1\over 2}(k_{12}+ k'_{12})
\quad \hbox{and} \quad
\tilde{k}_{23} \equiv {1\over 2}(k_{23}+ k'_{23}).
\end{equation}
Equation (\ref{pot2}) reduces to that of Eq.~(\ref{pot1})
with
\begin{eqnarray}
& k_1 = \lambda_1+\lambda_4, \qquad k_2=\lambda_2+\lambda_3+\lambda_4 &
\nonumber \\
& k_{12} = 2\lambda_4+\lambda_6+\lambda_7, \qquad k'_{12}=-\lambda_6 &
\nonumber \\
& k_{23} = 2\lambda_2-2\lambda_3+2\lambda_4+\lambda_5, \qquad
k'_{23}=-\lambda_5 & \nonumber \\
& k_4 = -2\lambda_7, \qquad k'_4=0.
\end{eqnarray}

\noindent
(a) The CP-even mass(-squared) matrix, $A_{ij} = A_{ji}$ in the basis
$(H_1, H_2, H_3)$ is given by
\begin{eqnarray}
&A_{11}  =  4k_1 u_1^2, \qquad
A_{12}  =  4\tilde{k}_{12} u_1u_2 + k_4 u_1u_3, & \nonumber \\
&A_{13}  =  4\tilde{k}_{12} u_1u_3 + k_4 u_1u_2, \qquad
A_{22}  =  4k_2 u_2^2 - {1\over 2}k_4 u_1^2(u_3/ u_2), & \nonumber \\
&A_{23}  =  4\tilde{k}_{23} u_2u_3 + {1\over 2}k_4 u_1^2, \qquad
A_{33}  =  4k_2 u_3^2 - {1\over 2}k_4 u_1^2(u_2/ u_3). &
\end{eqnarray}
In general, $A_{ij}$ has no zero eigenvalues. So there are three
real massive physical scalars.

\vskip 0.5cm
\noindent
(b) The CP-odd mass(-squared) matrix, $B_{ij} = B_{ji}$ in the basis
$(\eta_1, \eta_2, \eta_3)$ is given by
\begin{eqnarray}
&B_{11}  =  -2k_4 u_2u_3, \qquad
B_{12}  =  k_4 u_1u_3, &\nonumber \\
&B_{13}  =  k_4 u_1u_2, \qquad
B_{22}  =  - {1\over 2}k_4 u_1^2(u_3/ u_2), & \nonumber \\
&B_{23}  =  - {1\over 2}k_4 u_1^2, \qquad
B_{33}  =  - {1\over 2}k_4 u_1^2(u_2/ u_3). &
\end{eqnarray}
$B_{ij}$ has two zero eigenvalues and hence there are two
Goldstone bosons and one CP-odd real scalar, $\eta_{\rm Phys}$.
The Goldstone fields corresponding to $Z_1$ and $Z_2$ are
given respectively by
\begin{eqnarray}
G_{Z_1} & = & {g_2\over m_{Z_1}\sqrt{2} c_W}
\left\{u_1\cos\phi\eta_1 + u_2(\cos\phi+4a s_W\sin\phi)\eta_2
 + u_3(\cos\phi-4a s_W\sin\phi)\eta_3 \right\} \nonumber\\
G_{Z_2} & = & {g_2\over m_{Z_2}\sqrt{2} c_W}
\left\{u_1\sin\phi\eta_1 + u_2(\sin\phi-4a s_W\cos\phi)\eta_2
 + u_3(\sin\phi+4a s_W\cos\phi)\eta_3 \right\} \nonumber\\
 & &
\end{eqnarray}
The physical CP-odd field is given by
\begin{equation}
\eta_{\rm Phys} = {2u_2u_3\eta_1 - u_1u_3\eta_2 - u_1u_2\eta_3
\over\sqrt{u_1^2(u_2^2+u_3^2)+4u_2^2u_3^2}}
\end{equation}
with its mass given by
$-{1\over 2}k_4\left[u_1^2\left({u_2/u_3}+{u_3/ u_2}\right)
+4u_2u_3\right] $.

\vskip 0.5cm
\noindent
(c) The charged scalar mass(-squared) matrix, $C_{ij} = C_{ji}$ in the
basis $(\phi^\pm_1, \phi^\pm_2, \phi^\pm_3)$ is given by
\begin{eqnarray}
C_{11} & = & -k'_{12}(u_2^2+u_3^2)-k_4 u_2u_3 \nonumber \\
C_{12} & = & k'_{12} u_1u_2 + {1\over 2}k_4 u_1u_3 \nonumber \\
C_{13} & = & k'_{12} u_1u_3 + {1\over 2}k_4 u_1u_2 \nonumber \\
C_{22} & = & -k'_{23} u_3^2
- \left(k'_{12}+{1\over 2}k_4 {u_3\over u_2}\right)u_1^2 \nonumber \\
C_{23} & = & k'_{23} u_2u_3 \nonumber \\
C_{33} & = & -k'_{23} u_2^2
- \left(k'_{12}+{1\over 2}k_4 {u_2\over u_3}\right)u_1^2
\end{eqnarray}
There are two massive charged scalars and one Goldstone boson
associated with the $W^\pm$ boson:
\begin{equation}
G^\pm = {g_2\over m_W\sqrt{2}}\left(
u_1\phi^\pm_1 + u_2\phi^\pm_2 + u_3\phi^\pm_3 \right).
\end{equation}

\newpage
\leftline{\bf Figure captions}
\vskip 1.0cm

\noindent
Fig. 1. (a) The one-loop contribution to $\Delta a_\mu$
and (b) the accompanying diagram with the Goldstone field,
$G_{Z'}$, in the $R_\zeta$ gauge.
\vskip 0.5cm

\noindent
Fig. 2.(a) and (b): Constraints on the $(m_{Z'}, |a|)$ parameter
space in the exact discrete symmetry model --
(i) gauge boson masses (the allowed region is the area above the solid line)
(ii) $\Delta a_\mu$ (area below the dashed line)
(iii) 3-body decay (area below the dot-dashed line)
(iv) 2-body decay (area below the dotted line).
\vskip 0.5cm

\noindent
Fig. 3.(a) and (b): The same as Fig. 2 but for the case of
spontaneously broken discrete symmetry.


\begin{thebibliography}{99}
\bibitem{footnote1}Note that these equations also give
\[
u_1^2 = {s_W^2 m_W^2\over 2\pi\alpha_{\rm em}}
\left\{1-{1\over 16 a^2\tan^2\theta_W}\left({m_{Z'}\over m_W}\right)^2
\right\}.
\]
For values of $a$ and $m_{Z'}$ close to saturating the bound of
Eq.~(\ref{bd}), $u_1$ becomes small so that it is possible for
certain quark Yukawa couplings, $\lambda_q = m_q / u_1$ (where $m_q$
is the generic quark mass), to become nonperturbative.
\bibitem{footnote2}In the broken discrete
symmetry case, there is an additional contribution to $\Delta a_\mu^{Z'}$
which could be potentially large. This comes from a graph similar to that
in Fig. (1) where the internal $\tau$-lines are replaced by $\mu$ ones.
This contribution is given by
\[
\Delta a_\mu^{Z'} \simeq - {5\alpha_{\rm em}\over 3\pi}
\left(m_\mu\over m_{Z'}\right)^2 {|a|^2\over c_W}\cos^2 2\theta_L,
\]
which is suppressed by the factor $\left(m_\mu\over m_{Z'}\right)^2$
for $m_{Z'} > m_\mu$. Ignoring this contribution will not affect
any of our conclusions since the two-body decay constraint is
so stringent for $m_{Z'} \leq 0.5$ GeV.

In this case also,
the previously forbidden decay $\tau \to 3\mu$ is now
allowed. However, the partial decay rate is proportional to $|a|^4$ and
is thus very small.
\end{thebibliography}
\end{document}